\documentclass[12pt]{article}
\input epsf.tex
\def\centeron#1#2{{\setbox0=\hbox{#1}\setbox1=\hbox{#2}\ifdim
\wd1>\wd0\kern.5\wd1\kern-.5\wd0\fi
\copy0\kern-.5\wd0\kern-.5\wd1\copy1\ifdim\wd0>\wd1
\kern.5\wd0\kern-.5\wd1\fi}}
\def\ltap{\;\centeron{\raise.35ex\hbox{$<$}}{\lower.65ex\hbox{$\sim$}}\;}
\def\gtap{\;\centeron{\raise.35ex\hbox{$>$}}{\lower.65ex\hbox{$\sim$}}\;}
\def\gsim{\mathrel{\gtap}}
\def\lsim{\mathrel{\ltap}}
\title{A Symmetry for the Cosmological Constant\vskip .3in}
\author{\large{David E. Kaplan\,\thanks{\tt dkaplan@pha.jhu.edu}
\mbox{  }and Raman Sundrum\,\thanks{\tt sundrum@pha.jhu.edu}}\ \ \\ \\
\emph{\small{Department of Physics and Astronomy}} \\ 
\emph{\small{Johns Hopkins University}} \\ 
\emph{\small{3400 North Charles St}}. \\ 
\emph{\small{Baltimore, MD 21218-2686}}}

\begin{document}
\baselineskip=17pt
\pagestyle{plain}
\begin{titlepage}
\vskip -.4in
\maketitle

\begin{abstract}
We study a symmetry, schematically
Energy $\rightarrow$ -- Energy, which suppresses matter contributions to the 
cosmological constant. The requisite negative energy fluctuations are identified 
with a ``ghost'' copy of the Standard Model.  Gravity explicitly, but weakly, violates 
the symmetry, and naturalness requires General Relativity to break down at 
short distances with testable consequences.  If this breakdown is accompanied 
by gravitational Lorentz-violation, the decay of flat spacetime by ghost production 
is acceptably slow. We show that inflation works in our scenario and can lead to 
the initial conditions required for standard Big Bang cosmology.
\end{abstract}
\thispagestyle{empty}
\setcounter{page}{0}
\end{titlepage}

\section{Introduction}

It is sometimes hoped that the mysteries of short-distance gravity
will help solve the notorious Cosmological Constant Problem 
\cite{Weinberg:1988cp}. 
However, it is difficult to see where the opportunity lies. The 
Feynman diagrams for matter renormalization of the cosmological constant 
involve only the couplings of long wavelength gravitational fields to 
the quantum Standard Model (SM), 
the domain in which General Relativity appears to work perfectly well. 
(See for example the discussion in Ref. \cite{Sundrum:2003jq}.)
Another hope has been to find a symmetry under which a small cosmological 
constant is natural. But the most obvious candidates, supersymmetry 
and conformal invariance, appear too badly broken in Nature to serve this purpose.
In this paper, we study a scenario
in which both hopes may be realized.  We employ a  discrete symmetry 
to suppress the cosmological constant, but one which leads to 
instabilities of flat spacetime via gravitational processes. 
Adequate suppression of these processes requires a drastic 
breakdown of General Relativity at shorter distances.

The discrete symmetry, described in the next section, leads to an effective 
Lagrangian essentially the same as that proposed in Ref. \cite{Linde:1984ir},
\begin{eqnarray}
\label{L}
{\cal L} &=& \sqrt{-g} \{M_{Pl}^2 R - \rho_0 \nonumber \\
&+& {\cal L}_{matt}(\psi, D_{\mu}) - 
{\cal L}_{matt}(\hat{\psi}, D_{\mu}) + \ldots \},
\end{eqnarray}
where $g_{\mu \nu}$ is the metric, $\psi$ denotes a set of  ``visible sector''
fields, including the SM, and $\hat{\psi}$ denotes an identical copy of the 
visible fields.\footnote{A later proposal in a similar vein
\cite{Linde:1988ws,Linde:2002gj} introduced a symmetry under which fields
are exchanged between two independent spacetimes.  However, the extension of the theory
to the quantum regime appears to be quite problematic.}
The matter Lagrangian {\it function}, ${\cal L}_{matt}$, 
is the same in both terms in which it appears, 
but with different arguments. We refer to the $\hat{\psi}$ as the 
``ghost sector'' because of the ``wrong'' sign in front of its
Lagrangian, including kinetic terms. 
Note that the $\psi$ and the $\hat{\psi}$ include 
separate sets of gauge fields and separate sets of charged matter. 
We will discuss the ellipsis in the next section.
The central point of Eq. (\ref{L})
is that the visible and ghost sectors have equal and opposite 
vacuum energies,  canceling in their contribution
 to the cosmological constant, leaving only the bare (and 
possibly small) constant, $\rho_0$. This idea has an obvious shortcoming, 
namely instabilities originating from the ghost sector. 
In this paper, we nevertheless take the idea seriously by identifying the 
underlying symmetry, analyzing quantum effects and deducing
the features required to make it a controlled and realistic scenario.  

Because of negative energy excitations in the ghost sector, the Minkowski
``vacuum'' cannot be the ground state. However, as long as  positive and 
negative energy fluctuations are completely decoupled, the 
Minkowski vacuum is stable. But with any coupling between the two, the vacuum 
can spontaneously decay into combinations of positive and negative energy 
states. Since kinematics alone do not prevent arbitrarily large mass 
particles being produced in such processes, even 
effective field theory breaks down, becoming useful 
only to the extent that the coupling between positive and negative 
energy fluctuations is very weak.  See Ref. \cite{Carroll:2003st,Cline:2003gs} 
for an earlier discussion of vacuum decays in the presence of ghosts.

In the presence of gravity all excitations of matter are 
necessarily coupled.  The need to prevent 
excessively rapid decay of the vacuum is one reason General Relativity, 
indeed gravitational Lorentz invariance itself \cite{Cline:2003gs}, 
 must break down at short distances. For  reviews of the
subject of Lorentz violation see Refs. \cite{LIV}. 
The initial conditions for successful Big Bang cosmology 
require an essentially empty ghost sector, which
can arise from an early inflationary phase \cite{Linde:1984ir}.

Quantum gravity  gives corrections to the perfect cancellation 
of vacuum energies of visible matter and ghosts.
Adequate suppression of these contributions to make the observed 
cosmological constant natural
very likely requires a breakdown in the gravitational force close 
to present experimental limits from sub-millimeter tests of Newton's 
Law \cite{Hoyle:2004cw,Adelberger:2003zx,Long:2003dx,Chiaverini:2002cb}. 
Future tests should be able to probe this 
breakdown. The possible connection between the resolution of the
 cosmological constant 
problem and the sub-millimeter scale was first 
made in Ref. \cite{Banks:1988je}.
Refs \cite{Sundrum:1997js,Sundrum:2003jq,Sundrum:2003tb} 
proposed that this resolution is realized by de-localizing 
the gravitational interaction with matter on this scale, in a manner 
consistent with the equivalence principle. The technical connection between
this proposal and the present paper will be discussed in future work \cite{us}.

This paper is organized as follows. In Section 2, we motivate our 
effective Lagrangian from the viewpoint of a visible/ghost matter 
discrete symmetry called
 ``energy-parity'', explicitly (but weakly) broken by gravitational dynamics. 
In Section 
3, we study quantum 
dynamics of matter in a fixed gravitational background and show that 
there is no instability from negative energies at this level, and that the matter 
contributions to the cosmological constant naturally cancel due to 
energy-parity. In Section 4, we estimate the quantum gravitational corrections 
to the cosmological constant and (assuming naturalness)
use them to bound the cutoff scale on General Relativity. 
In Section 5, we show that as long as gravitational Lorentz violation occurs
at not much shorter distances than the cutoff of General Relativity, 
the instability of flat spacetime due to the negative energy fluctuations is consistent
with observation. In Section 6, we discuss the classical laws of 
gravity in the presence of negative energy fluctuations. 
In Section 7, we discuss our symmetry 
mechanism for controlling the cosmological constant when there are metastable 
vacua in the matter sector. In Section 8, we show how inflation 
can naturally lead to the cosmological initial conditions needed 
for our scenario. Finally, Section 9 discusses our results.

\section{Energy-Parity}

In order to motivate Eq. (\ref{L}) from a symmetry point of view, we 
begin by neglecting gravity and formally consider
 a ${\bf Z}_2$ ``energy-parity'' symmetry operation $P$, with $P^2 = 1$, 
acting on the matter Hilbert space.  However, instead of commuting with the 
Hamiltonian, $H$, like a standard symmetry operator, energy-parity satisfies
\begin{equation}
\label{parity}
\{ H, P \} \equiv HP + PH = 0.
\end{equation}
Thus, an energy eigenstate, 
\begin{equation}
H |E \rangle = E |E \rangle, 
\end{equation}
is transformed into one with the {\it opposite} energy,
\begin{equation}
H P |E \rangle = - E P |E \rangle, 
\end{equation}
rather than a state degenerate with $|E \rangle$, as is the case for 
standard symmetries. 
We will implement this parity so a 
Poincar\'{e}-invariant state exists which is also energy-parity 
invariant, namely $P |0 \rangle 
= |0 \rangle$.  From this follows
\begin{equation}
\langle 0| \{H,P \} |0 \rangle = 2 \langle 0| H |0 \rangle = 0.
\end{equation}
This corresponds to a vanishing cosmological constant contribution when 
gravity is turned back on.

Our fields transform under energy-parity in the following way:
\begin{eqnarray}
\label{symmetry}
g_{\mu \nu}(x) &\rightarrow& g_{\mu \nu}(x) \nonumber \\
\psi(x) &\leftrightarrow& \hat{\psi}(x).
\end{eqnarray}
Naively, it would appear that 
the pure gravity sector respects 
energy parity in Eq. (\ref{L}), 
while the matter Lagrangian maximally violates it. 
However, the opposite is true. To see this, ignore
gravity and note that
 Eq. (\ref{symmetry}) must be accompanied by 
\begin{equation}
H \rightarrow - H, 
\end{equation}
in order to satisfy Eq. (\ref{parity}). Relating the Hamiltonian to the 
Lagrangian, 
\begin{eqnarray}
L &=& \int d^3 \vec{x} (\Pi \dot{\psi} + \hat{\Pi} \dot{\hat{\psi}}) - H
\nonumber \\
&=& \int d^3 \vec{x} (\Pi \frac{\delta H}{\delta \Pi} + \hat{\Pi}  
\frac{\delta H}{\delta \hat{\Pi}}) -  H,
\end{eqnarray}
we see that the 
Lagrangian and action should be {\it odd} under Eq. (\ref{symmetry}), in order 
to respect energy-parity.  Our 
matter action respects energy-parity, and the gravity action 
maximally and explicitly violates it.

Energy-parity alone does not preclude direct matter couplings between $\psi$ 
and $\hat{\psi}$, considered part of the ellipsis of Eq. (\ref{L}).
Such visible-ghost couplings must be present at some level 
since they receive contributions induced 
by quantum gravity loops.  These couplings, if present, would contribute 
to the decay of the vacuum.  However, if we assume these couplings 
have their minimal natural strength, they do not dominate any of our 
vacuum decay estimates in Section 5 and
we will thus ignore them. The remaining terms in the 
ellipsis of Eq. (\ref{L}) are  purely gravitational higher derivative 
terms. Again, their effects do not dominate any estimates in this paper.

\section{Fixed Gravitational Background}

Here, we study matter dynamics in a {\it fixed} soft (low-curvature) 
gravitational 
background.  At the purely classical level, 
the negative sign in front of the ghost sector Lagrangian poses no problem, because 
classically the sign of the  Lagrangian is physically irrelevant and the 
two sectors are completely decoupled without dynamical gravity. 
Assuming all neutrinos have mass (entirely for simplicity of 
exposition), 
the effective theory in the far 
infrared is
\begin{eqnarray} 
{\cal L}_{eff} &=& \sqrt{-g} \{ - \frac{1}{4} F_{\mu \nu}^2 - \rho_{vis} + 
\frac{1}{4} \hat{F}_{\mu \nu}^2 + \rho_{vis} \} \nonumber \\
&=& \sqrt{-g} \{ - \frac{1}{4} F_{\mu \nu}^2  + 
\frac{1}{4} \hat{F}_{\mu \nu}^2\}.
\end{eqnarray}
The matter vacuum energy contributions cancel between the visible and 
ghost sectors because of energy-parity.

Now consider the same situation in quantum field theory. 
As a warm-up  we simplify to the
 case where 
$\psi$ denotes a single real scalar field 
rather than the entire visible sector. 
Similarly, $\hat{\psi}$ denotes a single ghost 
scalar. Further, we simplify the gravitational background to be 
exactly Minkowski space, 
$g_{\mu \nu} = \eta_{\mu \nu}$. The leading matter Lagrangian can 
then be written,
\begin{equation}
{\cal L} = \frac{1}{2} (\partial_{\mu} \psi)^2 - \frac{1}{2} m^2 \psi^2 - 
\lambda \psi^4  
- \frac{1}{2} (\partial_{\mu} \hat{\psi})^2 + \frac{1}{2} m^2 \hat{\psi}^2
+ \lambda \hat{\psi}^4 ,
\end{equation}
and the quantized Hamiltonian density is given by
\begin{equation}
\label{toy}
{\cal H} = \frac{1}{2} \Pi^2 + \frac{1}{2} (\bigtriangledown
 \psi)^2 + \frac{1}{2} 
m^2 \psi^2 + \lambda \psi^4 - 
\frac{1}{2} \hat{\Pi}^2 - \frac{1}{2} (\bigtriangledown
 \hat{\psi})^2 - 
\frac{1}{2} m^2 \hat{\psi}^2 - \lambda \hat{\psi}^4.
\end{equation}
This is the sum of two decoupled Hamiltonians.  We can quantize both 
sub-sectors, with positive energies propagating forwards in time for  $\psi$ 
and negative energies propagating forwards in time for $\hat{\psi}$.
Even corrected by interactions, the zero-point energies of $\psi$ and $\hat{\psi}$
cancel, leaving zero net vacuum energy.  
Again, because the two sectors are completely decoupled, no pathology exists in the 
negative energy sector. From the viewpoint of that sector, we have 
merely renamed Energy (${\cal H}$) by $-$ Energy ($-$ ${\cal H}$).

Finally, let us consider our real case of interest, general quantum visible
matter in a soft gravitational background. Here, it is easier to use 
path integral methods. Because the dynamics in the two matter sectors are 
decoupled in the absence of gravitational dynamics, 
the partition functional factorizes,
\begin{equation}
{\cal Z} = \left(\int {\cal D} \psi ~ e^{i \int \sqrt{-g}\,
{\cal L}_{matt}(\psi, D_{\mu})}\right)
\left(\int {\cal D} \hat{\psi} ~ e^{- i \int \sqrt{-g} \,
{\cal L}_{matt}(\hat\psi, D_{\mu})}\right),
\end{equation}
where now $\psi$ and $\hat\psi$ are a generic set of interacting fields.
The opposite sign of  the ghost Lagrangian now appears as the 
the replacement $i \rightarrow -i$ in the path integral phase factor. 
Since we want to propagate positive energies forward in time in the 
visible sector, we 
choose the usual $``+ i \epsilon"$ prescription, while propagating negative 
energies forward in time in the ghost sector requires a $``- i \epsilon"$ 
prescription. All other factors of ``$i$'' in effective quantum 
field theory can be eliminated from the 
Feynman rules in position space by working exclusively 
in terms of real fields and couplings (taking real and imaginary 
components of any complex fields)\footnote{There is a subtlety in the 
case of fermions. Here we can work in terms of real Grassman fields, 
$\chi$.   However, because of their anti-commuting nature, bilinears made 
from them must be multiplied by ``$i$'' to be Hermitian, 
$(i \chi_1 \chi_2)^{\dagger} = i \chi_1 \chi_2$. Thus, terms in the Lagrangian 
$\sim \chi^{4n+2}$ will have a factor of ``$i$'' that cannot be eliminated. 
However, as long as there is a conserved fermion number, or fermion number
is violated only by $4n$-fermion operators (for example, the Standard Model or the Standard
Model with Dirac neutrinos), this extra 
``$i$'' translates into an overall pre-factor for 
 amplitudes  involving odd fermion number processes. 
Since these do not interfere with 
amplitudes for even fermion number, the extra ``$i$'' falls out of all 
physical probabilities. Thus, without changing the physics, one can 
make the replacement $i \rightarrow -i$ in front of ghost fermion 
bilinears relative to visible bilinears.}. 
Thus, the two matter sectors have identical position space
Feynman rules except for the replacement $i \rightarrow - i$ everywhere. 
If we integrate out some high energy physics (symmetrically) from 
each of the matter sub-sectors, we must get a local
 effective theory of the form,
\begin{equation}
{\cal Z}_{eff} = 
\left(\int {\cal D} \psi_{\rm IR} ~ e^{i \int \sqrt{-g} 
{\cal L}_{eff}(\psi_{\rm IR})}\right)
\left(\int {\cal D} \hat{\psi}_{\rm IR} ~ e^{- i \int \sqrt{-g} 
{\cal L}_{eff}(\hat{\psi}_{\rm IR})}\right),
\end{equation}
that is, the ghost and visible factors identical  except for
 the replacements $\psi \leftrightarrow \hat{\psi}, ~ i \rightarrow -i$. 
This demonstrates the matter-renormalization stability of both energy-parity
and the decoupling of the ghost and visible sectors.

If we integrate out all massive matter,
we arrive at an effective theory of just the photons coupled to the gravitational background,
\begin{eqnarray}
\label{Zeff}
{\cal Z}_{eff} &=& 
\left(\int {\cal D} A_{\mu} ~ e^{i \int \sqrt{-g} (- F^2/4 - \rho_{vis})}\right)
\left(\int {\cal D} \hat{A}_\mu ~ e^{- i \int \sqrt{-g} (- \hat{F}^2/4 - 
\rho_{vis})}\right) 
\nonumber \\
&=&
\left(\int {\cal D} A_{\mu} ~ e^{i \int \sqrt{-g}  (-  F^2/4)}\right)
\left(\int {\cal D} \hat{A}_\mu ~ e^{- i \int \sqrt{-g} (- \hat{F}^2/4 )}\right).
\end{eqnarray}
Energy-parity forces the cancellation of the 
cosmological constant
induced by  the quantum visible sector, $\rho_{vis}$, 
against the corresponding term induced by the ghost sector.

\section{Quantum Gravity}

We next need to consider a 
cutoff, $\mu$, on graviton momenta, 
below which we trust Eq. (\ref{L}). We will find that 
in order to adequately suppress gravitational violation of energy-parity 
as well as vacuum decay, $\mu$ 
must be much smaller than the weak scale energies to which we have tested the 
SM. Physically, $\mu$  represents the scale of 
unspecified new gravitational 
physics which serves to cut off amplitudes derived 
from Eq. (\ref{L}). We will further 
require this physics to be Lorentz-violating.
It is unorthodox to contemplate the 
breakdown of General Relativity at energies below the breakdown of SM 
quantum field theory, or to consider a fundamental breakdown of Lorentz 
invariance at any scale, 
but these ingredients are central to our plot, and
we are unaware of any rigorous objections.

Because of energy-parity in the matter sector,
the only corrections to the cosmological term are induced by 
a variety of
 quantum gravitational corrections. 
 The gravitational sector itself naturally induces a quantum vacuum energy 
 of order $\mu^4$. It is therefore technically natural for the bare 
cosmological constant in Eq. (\ref{L}) to be of this same order, 
\begin{equation}
\rho_0 \sim {\cal O}(\mu^4).
\end{equation}
From now on we will assume this to be the case. 
Given the observed dark energy 
$\sim (2 \times 10^{-3} {\rm eV})^4$ \cite{Riess:1998cb} 
\cite{Perlmutter:1998np} \cite{cosmo}, 
naturalness implies 
\begin{equation}
\label{mu}
\mu \lsim 2 \times 10^{-3} {\rm eV}. 
\end{equation}
This corresponds to a length scale, $1/\mu \sim 100$ microns. A 
more refined estimate, including factors of $(4 \pi)$, 
gives a minimal breakdown length of $30$ microns \cite{Sundrum:2003jq}. Such a 
breakdown of General Relativity should  be probed by ongoing 
sub-millimeter tests of Newton's Law which have, so far, probed gravity down 
to $200$ microns  \cite{Hoyle:2004cw}.

\begin{figure}[htb]
\vskip 0.0truein
\centerline{\epsfysize=3.6in
{\epsffile{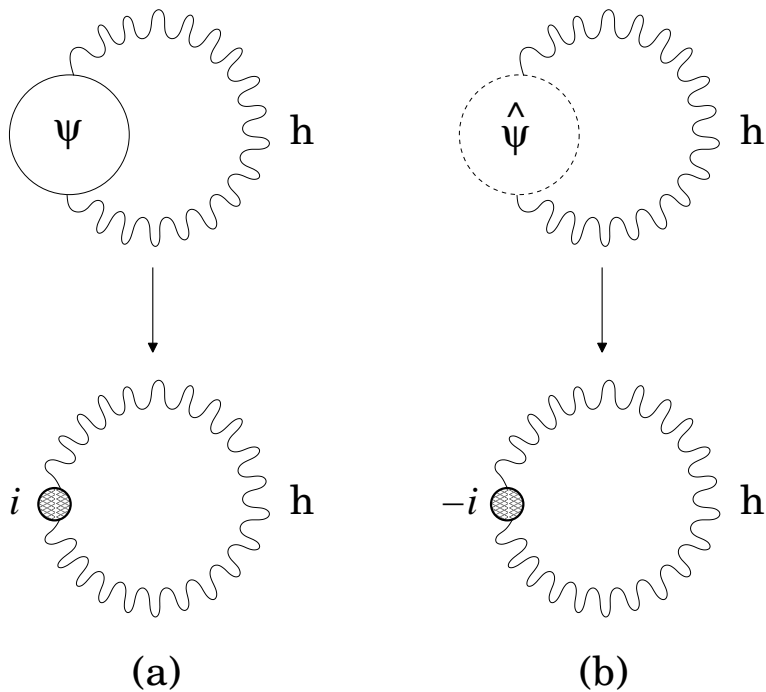}}}
\vskip 0.0truein
\caption[]{Hard matter (a) and ghost (b) loops attached to graviton loops and their effective vertices. }
\label{fig:gravityloop}
\end{figure}

Quantum gravity corrections to matter vacuum energy do not cancel completely, 
but the largest contributions, from scales above $\mu$, do cancel between 
the two matter sectors. To see this, note that since
graviton momenta are cut off at $\mu$, we can imagine integrating out 
all matter physics harder than $\mu$ before integrating out gravitons.
This must generate local effective 
vertices for the gravity (see Figure \ref{fig:gravityloop}). 
As discussed in Section 3 in path integral 
terms, the visible and ghost contributions are related by $i \rightarrow -i$. 
For a local vertex, the only factor of ``$i$'' is the pre-factor of the 
action in the path integral. Thus, these hard matter contributions must 
cancel between the two matter sectors. 
This leaves integrating out gravitons, cut off by $\mu$, as well as 
 matter physics
softer than $\mu$, say from photon loops and ghost-photon loops.
These do not generally cancel, because the non-local 
gravity effective action not only has a pre-factor of $i$, but also 
has imaginary parts from soft matter cuts.  
By dimensional analysis the leading  
contributions of this type to the cosmological constant are
\begin{equation}
\label{mu2}
\delta \rho \sim \mu^6/M_{Pl}^2.
\end{equation}  
For $\mu \sim {\cal O}(10^{-3} {\rm eV})$, these contributions are negligible.

There is a caveat in our prediction for the breakdown of 
Newton's Law in sub-millimeter tests.  Inferring the value of the cosmological
constant from observations followed by 
requiring naturalness in the pure quantum gravity sector contribution to the
cosmological constant (as above), yields the most accessible length scale for the 
breakdown of General Relativity.  Indeed,
we will show elsewhere \cite{us} that these estimates do not change even if 
one begins with a supersymmetric gravitational sector, given that the 
SM is not supersymmetric below a TeV.  However, it is in principle possible that 
the gravitational physics that acts to cut off these contributions does not
couple appreciably to SM matter, and therefore would not show up in 
sub-millimeter tests. In this case, there is still a 
prediction following from the contributions of Eq. (\ref{mu2}), which 
rests robustly 
on the coupling of the gravity cutoff physics to matter. Requiring $\delta\rho$
in Eq. (\ref{mu2}) be on the order of
the observed dark energy 
yields a prediction for the breakdown of Newton's Law at 
distances
\begin{equation}
1/\mu \sim 1/{\rm (10 MeV)} \sim 10\, {\rm fm}.
\end{equation}

\section{Vacuum Decay}

We now consider the inevitable 
instability implied by the full dynamics of the ghost sector coupled to gravity. 
We assume that in the far past, the ghost sector starts off 
close to empty, while the visible sector 
 and gravity are close to their state in standard 
cosmology.  The question arises how rapidly 
physical processes can exploit the negative energy states of the 
ghost sector in order to populate both that sector as well as the visible and 
gravity sectors. A similar situation, in the context of ``phantom'' dark energy
\cite{phantom} was analysed in Refs. \cite{Carroll:2003st,Cline:2003gs}. 

To see what issues are involved, let us first 
return to our toy 
example of Eq. (\ref{toy}), now adding a perturbation connecting the 
regular and ghost-like sectors,
\begin{equation}
\delta {\cal H} = g \psi^2 \hat{\psi}^2. 
\end{equation}
Such a vertex  destabilizes the 
 Minkowski vacuum (empty space) by allowing energy conserving processes 
such as (Nothing) $\rightarrow \psi(k_1) + \psi(k_2) + \hat{\psi}(p_1) 
+ \hat{\psi}(p_2)$. At leading order for this process, the event rate per unit 
time per unit volume is
\begin{eqnarray}
{\cal P}_{\rightarrow \psi\psi\hat{\psi}\hat{\psi}}  &\sim & g^2 \int d^4 p_1 \int d^4 p_2 \int d^4 k_1 \int d^4 k_2 
\delta(p_1^2 - m^2) \delta(p_2^2 - m^2) \delta(k_1^2 - m^2) 
\delta(k_2^2 - m^2) \nonumber \\ 
&\times & \theta(-p_{10}) \theta(-p_{20}) \theta(k_{10})
\theta(k_{20}) \delta^4(p_1 + p_2 + k_1 + k_2).
\end{eqnarray}
This type of calculation resembles ordinary $2 \rightarrow 2$ 
cross-section calculations for $(-p_1) + (-p_2) \rightarrow k_1 + k_2$.
However, while in $2 \rightarrow 2$ scattering the initial momenta, 
$p_1, p_2$ are {\it given} and we only must integrate over the final 
phase space of $k_1, k_2$, for vacuum decay we must obviously also 
integrate over the phase space for $p_1, p_2$. We will massage this extra 
phase space integration by defining $P \equiv p_1 + p_2, ~ p \equiv 
p_1 - p_2$, and insert the identity (since $P$ is always time-like or null), 
\begin{equation}
\int_0^{\infty} ds \delta(P^2 - s) = 1.
\end{equation}
Further noting that the on-shell $\delta$-functions for the $p_i$ 
satisfy, 
\begin{equation}
\delta((P+p)^2 - m^2) \delta((P-p)^2 - m^2) \propto 
\delta(p^2 + P^2 - m^2) \delta(P.p), 
\end{equation}
we find
\begin{eqnarray}
{\cal P}_{\rightarrow \psi\psi\hat{\psi}\hat{\psi}}  &\sim & 
g^2 \int_0^{\infty} ds \int d^4 P \delta (P^2 - s) \theta(-P_0)  
\int d^4 p \theta(-p_0) \delta(p^2 + s - m^2) \delta(P.p) \nonumber \\
&\times & \int  \frac{d^3 \vec{k}_1}{2 \omega_{k_1}} 
\int \frac{d^3 \vec{k}_2}{2 \omega_{k_2}}
 \delta^4(k_1 + k_2 + P).
\end{eqnarray}
Finally, defining $v_{\mu} \equiv - P_{\mu}/\sqrt{s} $, we arrive at the 
simple form,
\begin{eqnarray}
{\cal P}_{\rightarrow \psi\psi\hat{\psi}\hat{\psi}}  &\sim &  
g^2 \int \frac{d^3 \vec{v}}{2 \sqrt{1 + \vec{v}^2}} 
\int_0^{\infty} ds ~ \sqrt{s}  
\int d^4 p \theta(-p_0) \delta(p^2 + s - m^2) \delta(-p.v) \nonumber \\
&\times&
\int  \frac{d^3 \vec{k}_1}{2 \omega_{k_1}} 
\int  \frac{d^3 \vec{k}_2}{2 \omega_{k_2}}
 \delta^4(k_1 + k_2 - \sqrt{s} v),
\end{eqnarray}
where $v_{\mu}$ has become a $4$-velocity with $v_0 \equiv \sqrt{1 + \vec{v}^2}$.

Notice that for {\it fixed} $v$ and $s$, the remaining phase space 
integrals over $p, \vec{k}_1$ and $\vec{k}_2$ are necessarily finite. In 
particular they give some Lorentz-invariant function of $s$ and $v_{\mu}$ -- 
really a function of $s$ alone since $v^2 = 1$. For $s \gg m^2$, this 
function scales as $\sqrt{s}$ by dimensional analysis. Thus, the integral 
over $s$ has a serious power divergence even in this tree level calculation. 
Assume that some new Lorentz-invariant physics appears at a scale 
$s_{max}$ to cut off this divergence and give a finite $s$ 
integral.  This still leaves the integral over $\vec{v}$, which 
diverges quadratically because of the $v$-independence deduced above. 
There is no way for any new Lorentz-invariant physics 
to cut off this divergence in the decay probability. To get a 
finite answer we must have  Lorentz-violating  physics act as the cutoff 
\cite{Cline:2003gs} at 
a scale ${\cal E}$, 
 and assume Lorentz-invariance is a (very good)
approximate symmetry below this scale. 
Thus, in our toy example, we estimate 
\begin{equation}
{\cal P}_{\rightarrow \psi\psi\hat{\psi}\hat{\psi}}   \sim g^2 {\cal E}^2 
s_{max}.
\end{equation}

The decomposition of the phase space integrals for the decays of the vacuum 
seen in the above example generalize straightforwardly to all varieties of 
such processes at any order in perturbation theory. There is an overall 
integral over the total ghost $4$-momentum, 
\begin{equation}
\int d^4 P ... = \int d^3 \vec{v}/2 \sqrt{1 + \vec{v}^2} \int ds s ...,
\end{equation}
with all phase space integrals over relative ghost momenta and 
all visible momenta yielding some finite function of $s$ alone if all 
physics is Lorentz-invariant. New Lorentz-invariant physics 
can cut off the $s$ integral, but there is always an overall divergent 
$\int d^3 \vec{v}/2 \sqrt{1 + \vec{v}^2}$ which can only be cut off by 
invoking high-energy Lorentz-violation. Since, in our scenario, 
such processes connecting the ghost and other sectors always go via the 
gravitational sector, we will identify both  $\sqrt{s_{max}}$ 
and ${\cal E}$ with 
$\mu$.\footnote{
In principle, we  could simply consider two gravitational cutoffs, 
$\sqrt{s_{max}} 
\neq {\cal E}$, which 
 would change some of our estimates. We have adopted a single 
gravitational cutoff scale, $\mu$, as the simplest option in this paper.
We discuss the implications of the alternative option in future work 
\cite{us}.}

The dominant two processes for vacuum decay are 
\begin{eqnarray}
{\rm Nothing} &\rightarrow& \psi + \psi + \hat{\psi} + 
\hat{\psi}, \nonumber \\
{\rm Nothing} &\rightarrow& ({\rm graviton\! - \! excitation}) + \hat{\psi} + 
\hat{\psi} .
\end{eqnarray}
The first process is mediated by an off-shell graviton, so its decay rate is suppressed by $1/M_{Pl}^4$.  The second process includes the possibility of producing excited gravitons (or other particles in the gravitational sector) responsible for cutting off gravity at the scale $\mu$.  The first process is the dominant production of visible matter from the vacuum.  For the case of massless $\psi$ (i.e., photons), we use dimensional analysis, and include the (4$\pi$)s which result from phase space integrals to estimate
\begin{eqnarray}
{\cal P}_{\rightarrow \gamma \gamma \hat\gamma \hat\gamma} &\sim &
\frac{1}{4\pi}\left(\frac{1}{8\pi^2}\right)^2\frac{\mu^8}{M_{Pl}^4}\nonumber\\
&\sim& 2\times 10^{-92} \left(\frac{{\mu}}{2\times 10^{-3} {\rm eV}}\right)^8  [{\rm cm}^3 \times 10{\rm Gyr}]^{-1}. 
\end{eqnarray}
Given our estimate of $\mu$, the number of photons produced over the lifetime of the universe is completely negligible.  We become experimentally sensitive 
to this process in the cosmic ray background \cite{MeV-gammas}
only when $\mu \gsim $ MeV  \cite{Cline:2003gs}. 
This may become relevant
if the possibility of a higher gravitational cutoff, as discussed at the end of the last
section, is realized.

The second process, the decay into excited gravitons and ghosts, is the dominant process populating the universe with ghosts.  We take 
the gravity-sector particles to have mass $\sim\mu$ and their coupling to (ghost) matter of gravitational strength.  A rough estimate of the current energy density of ghost radiation due to this decay is the rate times the age of the universe times the average energy of the ghosts (which again will be of order $\mu$).  The estimate
\begin{eqnarray}
t_0 \; \mu \; {\cal P}_{\rightarrow h^* \hat\gamma \hat\gamma}  &\sim &
\frac{1}{4\pi}\left(\frac{1}{8\pi^2}\right)\frac{\mu^7}{M_{Pl}^2} t_0\nonumber\\
&\sim& (2\times 10^{-12} {\rm eV})^4 \left(\frac{{\mu}}{2\times 10^{-3} 
{\rm eV}}\right)^7  \left(\frac{t_0}{10{\rm Gyr}}\right) ,
\end{eqnarray}
is negligible compared with, for example, the radiation energy density today.

We have focused on photons as massive particles are even less important due to the inherent phase-space cutoff $\mu$ on all processes.

\section{Classical Gravity with Ghosts}

In a fixed 
 external gravitational field such as we considered 
in Section 3, 
ghost  matter behaves identically to ordinary matter,
the only distinction being the overall sign of the Lagrangian, which is 
irrelevant to the equations of motion. For example, if a ghost  
particle is brought near the Earth it will fall towards the ground. 
A ghost mass and a visible mass will be repelled from each other, however,
if the ghost mass dominates. This is because we can think of the 
ghost mass as setting up a gravitational field which is then felt by the 
smaller visible mass. Since the sign of the ghost stress tensor is reversed, 
the linearized gravitational field set up is also reversed. The visible 
mass sees this reversed field, and is repelled rather than attracted.
Thus, there must be a transition from attraction to repulsion depending on 
the masses. 
The gravitational force between two ghosts is also repulsive. 
To see this, note that 
since the overall sign of the action is irrelevant, it is 
simpler to think of the gravitational sector as being ghost-like relative to 
the ghost sector. The non-relativistic force law arises diagrammatically from 
one-graviton  exchange. The relative ghost-like nature of 
gravity implies a sign-flip for the graviton  compared 
to the usual computation.  In the ghost sector, therefore, 
the usual law of universal attraction is replaced by universal repulsion. 

All these results are  illustrated by writing  the Lagrangian 
in the non-relativistic approximation for the relative motion between two 
masses:
\begin{equation}
L_{relative} = \frac{1}{2} \frac{m_1\, m_2}{m_1 + m_2} \dot{\vec{r}}^2 
+ G_N \frac{m_1\, m_2}{|\vec{r}|}.
\end{equation}
The kinetic coefficient is just the usual formula for the reduced mass.
This formula continues to hold even in the presence of ghost masses, the only 
difference being that these ghost masses must be considered {\it negative}.
There are, of course, relativistic corrections to the static force, most 
significantly gravitational radiation from accelerating ghost masses. 
Instead of slowing such masses, the emission 
of gravitational radiation speeds them up.

The center-of-mass coordinate is cyclic and decouples from the relative motion
as usual.  However, in the limit $m_1 = - m_2 \equiv m$, the relative coordinate $\vec{r}$
becomes cyclic and the center of mass coordinate becomes proportional to $\vec{r}$.
The more useful coordinate is the average position $\vec{R} = \vec{x}_1 + \vec{x}_2$
with the equation of motion 
\begin{equation}
\ddot{\vec{R}} = \frac{2 G_N m}{r^2}\hat{r} ,
\end{equation}
in which case the matter-ghost system spontaneously accelerates in the direction $\hat{r}$, while their
relative positions remain fixed.

To avoid such an exotic type of dark matter (which would spoil 
standard cosmology were it to (co-)dominate), we require the ghost 
sector to be far more empty than the visible
 sector of our universe. This should be
considered a (plausible) 
requirement on the initial conditions of the universe.
As long 
as the negative energy density and pressure of ghost matter is subdominant, 
the expansion of the universe is driven by visible matter, and the cosmological 
term in standard fashion. 

There is a constraint on how far back in cosmological time
 our effective theory continues to make sense, following from the 
gravitational cutoff
 $\mu^2$  on spacetime curvature. From Einstein's Equations this corresponds 
to an energy density in the matter sector of order $\mu^2 M_{Pl}^2 \sim$ 
TeV$^4$. 
Thus we are constrained to cosmology from roughly just above the 
electroweak phase transition to the present.

\section{Metastable Matter Vacua}

\begin{figure}[htb]
\vskip 0.0truein
\centerline{\epsfysize=3.6in
{\epsffile{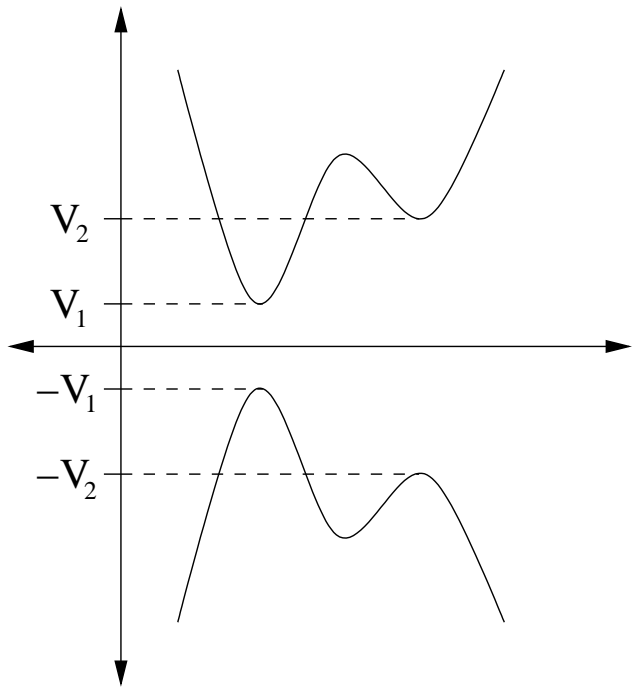}}}
\vskip 0.0truein
\caption[]{Meta-stable vacua in a theory with  energy parity. }
\label{fig:ccvac}
\end{figure}

In any proposal which claims to attack the cosmological constant problem,
one must consider what happens when the matter sector has a metastable vacuum
as well as a true vacuum, in order to understand what principle determines which 
vacuum has the suppressed cosmological constant.
In the present scenario, a metastable vacuum in the visible sector 
must be reflected in the ghost sector, as illustrated in Fig. \ref{fig:ccvac}. 
 The corresponding 
vacuum energies are taken to be $V_2 > V_1$ and $-V_2 < -V_1$.  When 
the two sectors occupy parity-symmetric 
vacua at $\pm V_1$ or $ \pm V_2$,  
the matter contribution to the cosmological constant vanishes (up to the small 
quantum gravitational corrections we estimated earlier). 
But if the visible sector is at $V_2$ while the ghost sector is 
at $-V_1$, the matter 
contribution to the cosmological constant is $V_2 - V_1 > 0$. Similarly, 
if the visible sector is at $V_1$ while the ghost sector is at $- V_2$, 
a negative cosmological constant, $V_1 - V_2$ emerges 
for the lifetime of the metastable state.  If the metastable vacua 
decay in the long run (before the vacuum energy becomes dominant),
then eventually both matter sectors will be near $\pm V_1$ 
where the matter contribution to the cosmological constant
vanishes.

\section{Inflation}
The desired initial conditions of our scenario -- standard big bang cosmology beginning before nucleosynthesis and an empty ghost sector -- is surprisingly easy to achieve using cosmic inflation.
Here we elaborate on Ref. \cite{Linde:1984ir}.  
The continuity equation for the ghost sector is the same as that for ordinary matter.  For
an approximately homogeneous and isotropic universe, the equation is
\begin{equation}
0= T^\mu_{\:\:\nu ; \mu} \simeq \frac{\partial\rho_{ghost}}{\partial t} + 3(\rho_{ghost} + p_{ghost}) \frac{\dot a}{a} , 
\label{eq:Econ}
\end{equation}
where $\rho_{ghost}$ and $p_{ghost}$ are the energy density and pressure of the ghost sector and $a$ is the scale factor in the Robertson-Walker metric.  While $\rho_{ghost}$ (and for radiation, $p_{ghost}$) is negative, the energy density clearly scales like that of normal matter and radiation, namely $\rho_{ghost} \sim a^{-4}$ for radiation and $\rho_{ghost}\sim a^{-3}$ for non-relativistic matter.  Thus, if the universe were in a state in which positive 
vacuum energy dominated, both normal radiation and matter and ghost radiation and matter would dissipate due to the exponentially growing scale factor.

\begin{figure}[htb]
\vskip 0.0truein
\centerline{\epsfysize=3.6in
{\epsffile{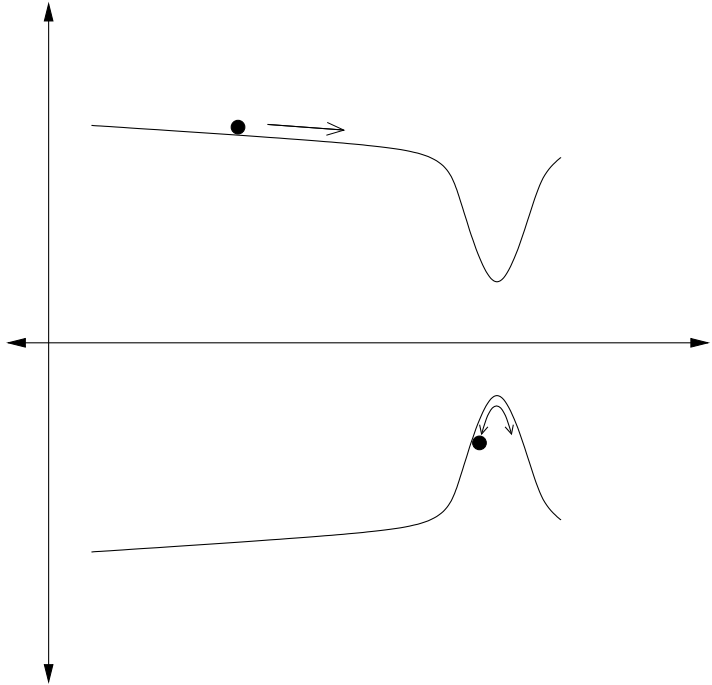}}}
\vskip 0.0truein
\caption[]{Inflation in a theory with energy parity. }
\label{fig:inflation}
\end{figure}

Inflation can be generated by the displacement of a scalar field from its minimum as long as its potential is flat enough (see Figure \ref{fig:inflation}).  One necessity is that the corresponding field in the ghost sector is sitting at its maximum for the number of e-foldings required by inflation.  In fact, the ghost inflaton could be initially displaced, so long as its displacement is smaller than that of the inflaton and positive vacuum energy dominates.  The ghost partner's dynamics will be governed by Eq. (\ref{eq:Econ}) with $\rho_{ghost} = -(\partial_t \hat\phi)^2 - (\nabla \hat\phi)^2/(2 a^2) - V(\hat\phi)$ and $p_{ghost} = -(\partial_t \phi)^2 + (\nabla \phi)^2/(2 a^2) + V(\hat\phi)$, leading to
\begin{equation}
\ddot{\hat{\phi}} + 3\, \frac{\dot a}{a}\, \dot{\hat{\phi}} + V'(\hat\phi) =0 ,
\end{equation}
where $V(\hat\phi)$ is the potential function in ${\cal L}_{matt}$, and is thus the same in the two sectors.  The ghost inflaton has the same equation of motion as the visible sector inflaton, but
because of the overall minus sign, the actual potential for the ghost inflaton is inverted compared with that of the visible sector.  The displaced ghost 
rolls {\it up} the potential towards its maximum, performs coherent oscillations and/or reheats into ghost radiation.  Positive vacuum energy dominates and the ghost radiation dilutes as the universe inflates.  Finally, the standard inflaton rolls towards its minimum, perhaps performs coherent oscillations, and then reheats an otherwise empty universe with only visible matter.  This symmetric looking evolution of the ghost inflaton is what one expects -- dynamical gravity breaks the energy parity symmetry, but coupling to the background metric does not.

In the simplest picture, the model-building challenge is only to construct an inflation scenario in which the curvature (and therefore the Hubble scale) during inflation is less than $\mu$, the energy-momentum cutoff of gravity.  Speculation about, for example, the physics responsible for the cutoff $\mu$ may lead to possibilities outside of inflation which address the question of initial conditions of our universe.

\section{Discussion}
We proposed a simple symmetry to control 
the cosmological constant, but one with an apparently fatal flaw, the instability 
of flat spacetime.  The danger from this instability, however, is entirely sensitive 
to features of short distance gravity outside the currently probed 
experimental regime.  This scenario, therefore, changes the 
character of the cosmological constant 
problem. The strictest bounds on the distances at which gravity 
must be modified in fact arise from the explicit breaking of the protective 
symmetry by gravity, putting such modifications within reach of ongoing tests
of short distance gravity \cite{Hoyle:2004cw,Adelberger:2003zx,Long:2003dx,Chiaverini:2002cb}. 
Controlling the instability does, however, introduce a new qualitative
 requirement, namely gravitational Lorentz-violation. 
This is (model-dependently) another source of potential experimental signals 
\cite{GLV}. 
Gravitational Lorentz 
violation can also radiatively induce Lorentz violation in visible matter, but 
these effects (in fractional shifts in maximal speeds) are negligibly 
 small, $\lsim {\cal O}(\mu^2/M_{Pl}^2) \sim 10^{-60}$.
Our scenario has an 
 acceptable cosmology provided we have initial conditions with 
 the ghost sector very sparsely populated, and we showed how inflation can 
 make these initial conditions natural. Ghost matter has unusual gravitational 
 laws and unusual equations of state. If enough of it has 
 survived to the present, it may provide interesting signals in precision 
 cosmological measurements.

It is of course important to understand how to build consistent 
theories with gravitational breakdown energy scales far below that of 
non-gravitational particle physics, as well as how to incorporate ghost 
matter at the fundamental level. For gravitational 
Lorentz violation, there are two possible scenarios. First, Lorentz invariance may 
not be a fundamental symmetry of Nature, but rather some sort of 
accidental or emergent symmetry. Of course this implies 
emergent General Relativity \cite{emergent} of some sort. 
Second, it may be that Lorentz invariance is a fundamental symmetry, but the gravitational 
vacuum spontaneously breaks this symmetry. An example is the 
effective field  theory of Ref. \cite{Arkani-Hamed:2003uy}. Gravitational 
fluctuations about such a vacuum need not be constrained by exact Lorentz 
invariance.  Exploration of the character of this Lorentz-violating cutoff may hold the 
key to additional experimental tests of our scenario.

\vskip.4in
We thank Markus Luty for useful comments, especially on Section 6, 
and Ann Nelson and Frank Petriello for 
helpful discussion and comments. We are grateful to Steve Hsu for 
an inspiring seminar on the work of Refs. \cite{Hsu:2004vr} connecting 
the null energy condition to instabilities as well as discussion.  We are 
also grateful to Andrei Linde for informing us of Ref. \cite{Linde:1984ir}
and for discussions on Refs. \cite{Linde:1988ws, Linde:2002gj}.
DK is supported in part by NSF grants P420-D36-2041-4350 and 
P420-D36-2043-4350, by the DOE's OJI program under grant DE-FG02-03ER41271, 
and by the Alfred P. Sloan Foundation.  RS is supported in part by NSF grant
P420-D36-2043-4350.

\end{document}